\documentclass[5p,times]{cas-dc}

\graphicspath{{figures/}}
\DeclareGraphicsExtensions{.pdf,.jpeg,.png}

\usepackage{enumitem}
\usepackage[cmex10]{amsmath}
\usepackage{array}
\usepackage{url}
\hypersetup{
    colorlinks=true,
    linkcolor=blue,
    urlcolor=blue,
    citecolor=blue
}
\usepackage{amsmath,amssymb}
\usepackage{caption}
\usepackage{subcaption}
\usepackage{float}
\usepackage{booktabs}
\usepackage{multirow}
\usepackage{makecell}
\usepackage{graphicx}
\usepackage{color}
\usepackage{pifont}

\usepackage[numbers,square]{natbib}
\usepackage{placeins}
\usepackage{cuted}

\usepackage{float}
\floatstyle{plain}
\restylefloat{figure}
\restylefloat{table}

\newcommand{\cmark}{\ding{51}}
\newcommand{\xmark}{\ding{55}}
\newcommand{\pmark}{$\triangle$}
\newcommand{\expref}[1]{\hyperref[sec:e#1]{E#1}}
\newcommand{\casestudyref}{\hyperref[sec:case-study]{Case Study}}
\begin{document}
\let\WriteBookmarks\relax
\renewcommand{\topfraction}{0.95}
\renewcommand{\dbltopfraction}{0.95}
\renewcommand{\textfraction}{0.05}
\renewcommand{\floatpagefraction}{0.7}
\renewcommand{\dblfloatpagefraction}{0.7}

\shorttitle{Darpan: A Digital Twin Framework for the Computing Continuum}

\shortauthors{Wang, Buyya}

\title [mode = title]{Darpan: A Digital Twin Framework for the Next-Generation Computing Continuum}

\author[1]{Zhiyu Wang}
\ead{zhiyuw5@unimelb.edu.au}

\affiliation[1]{organization={The Quantum Cloud Computing and Distributed Systems (qCLOUDS) Laboratory, School of Computing and Information Systems, The University of Melbourne},
    city={Melbourne},
    country={Australia}}

\author[1]{Rajkumar Buyya}
\ead{rbuyya@unimelb.edu.au}

\begin{abstract}
Computing-continuum applications distribute work across devices, edge systems, fog resources, and clouds. While a placement, scheduling, or recovery decision is being made, resource availability, network conditions, and application progress may change, so the decision can be invalid by the time it is executed. Existing runtimes enact predetermined decisions, whereas simulation tools compare alternatives in preconfigured environments; neither explores alternative outcomes directly from the current state of a running application. We propose a Digital Twin framework, called Darpan, that supports both physical and digital execution: the physical side runs real applications and continuously observes their runtime state, while the digital side maintains a continuously updated, executable virtual counterpart based on these physical observations. Darpan evaluates candidate decisions independently from the same starting point and returns the selected decision to the physical side, where its feasibility is validated against the latest physical state before actual execution. Across real Directed Acyclic Graph (DAG) workloads, Darpan predicts physical response time with a mean absolute error of 0.485~s and retains 93.1\% scale-out efficiency at 40 nodes. In the state-change trials, Darpan takes about 2~ms on average from state capture to rejection of an invalidated decision, stopping the request before data transfer and avoiding unnecessary physical execution overhead. Under the same physical-DAG budget, Darpan-generated experience enables a weaker Proximal Policy Optimization (PPO) scheduler to outperform two state-of-the-art Deep Reinforcement Learning (DRL) schedulers by 29.1\% and 20.2\%.
\end{abstract}


\maketitle

\section{Introduction}

Today's computing continuum has grown by extending cloud services outward to fog, edge, and device resources, allowing computation to be placed closer to data sources and users~\cite{bittencourt2025continuum}. This environment spanning Device, Edge, Fog, and Cloud resources increasingly supports long-running, latency-sensitive, and data-intensive applications, while resource availability, network conditions, task dependencies, and system load may continue to change during execution~\cite{wang2024drlis,wang2025tfddrl,sedghani2025placement,zhao2025socialwelfare,liu2026dataorchestration,ren2026joint}. A conventional runtime typically observes the current system state, selects one placement, scheduling, or recovery decision, and then executes that decision directly on physical resources. This single-path execution creates two problems: the state used to make a decision may change before the decision is actually executed, while trying multiple candidate decisions directly on the live infrastructure consumes resources, incurs additional data movement, and may interfere with running applications. These changes require the next-generation computing continuum not only to observe and control ongoing physical execution, but also to explore alternative decisions before they affect the real system.

A Digital Twin is a digital counterpart continuously associated with a specific physical system. It receives new state from the real system, keeps its digital representation updated as the physical counterpart evolves, and uses this representation for monitoring, prediction, and exploration of the possible outcomes of different operations~\cite{xia2026streamingdt}. The direct benefit of a Digital Twin is that it allows a system to evaluate the possible effects of different operations based on continuously updated real-world state before those operations are applied to the physical system. This capability is particularly important for the next-generation computing continuum, where applications must continuously perceive environmental changes and adapt their runtime decisions accordingly. For example, in healthcare, wearable and medical devices can continuously collect patient state, while distributed Edge and Cloud resources maintain the corresponding Digital Twin to support continuous monitoring and decision making~\cite{chen2025medical}; in intelligent transportation, vehicles, roadside units, and Edge infrastructure continuously generate vehicle, road, and network information, which can be incorporated into Digital Twins to support traffic control, task offloading, and resource allocation~\cite{cao2025vehicular}; and in smart agriculture, sensors and connected devices continuously collect crop, soil, and environmental conditions, allowing Digital Twins to evaluate irrigation, environmental-control, and resource-management strategies~\cite{yang2025adaptivefl}.

However, existing systems provide only part of this Physical--Digital execution process. Continuum runtimes can deploy real applications, collect runtime state, and execute resource-management decisions on real resources~\cite{deng2021fogbus2,wang2025reinfog,rosendo2020e2clab}, but they primarily support physical execution: the system selects a decision and then executes it on the real infrastructure. Simulation and emulation tools can compare different strategies in a digital environment~\cite{calheiros2011cloudsim,sonmez2018edgecloudsim,mahmud2022ifogsim2,mechalikh2021pureedgesim,massa2026eclypse}, but these virtual executions typically start from preconfigured environments rather than directly from the current state of an application that is already running. Existing Digital Twin approaches have addressed self-healing, data access, and resource orchestration in the computing continuum~\cite{qin2024dtnsurvey,saxena2025selfhealing,gao2026cost,li2026dtorchestration}, but a general execution model is still missing that connects continuously evolving physical execution with digital execution capable of exploring multiple alternatives from the current state and then returns the selected operation to the latest physical system for execution.

To address this gap, we propose Darpan\footnote{The name Darpan comes from the Sanskrit word for ``mirror'', reflecting the framework's role of maintaining a synchronized digital reflection of a running physical system.}, a Digital Twin framework for the computing continuum that supports both physical and digital execution. Darpan's physical side executes real applications and continuously reflects the current runtime state, while its Digital Twin remains synchronized with the real system and provides an executable environment for exploring alternative decisions. Darpan captures the current state at a given time together with performance models learned from physical execution and their version identifiers in a \emph{Snapshot}. Multiple candidate decisions can execute independently from the same Snapshot, allowing them to be compared from the same starting point without first modifying the real system. After an external decision maker selects an operation based on the results of these digital executions, Darpan returns the selected operation to the physical side and checks whether it remains feasible against the latest physical state. Only an operation that passes this check is executed on the physical infrastructure. Through this process, Darpan connects physical observation, digital exploration, and physical execution in a continuous closed loop.

We evaluate Twin accuracy, adaptation to changing conditions, decision support, recovery, execution-plane scale-out, runtime cost, and the effect of state changes between decision making and physical execution. Darpan predicts physical DAG response time with a mean absolute error of 0.485~s and adapts to changes in compute and network conditions using new physical observations. As the execution plane grows from 10 to 40 nodes, Darpan retains 93.1\% scale-out efficiency. In the state-change trials, the path from state capture to rejection of an invalidated placement takes about 2~ms on average and stops the request before artifact transfer begins. After rejection, capturing a fresh state and submitting a predefined backup placement incurs a mean control overhead of 4.066~ms for node unavailability and 1.628~ms for insufficient processor capacity. Under the same physical-DAG execution budget, Darpan-generated experience further enables the same PPO learner to outperform two state-of-the-art DRL schedulers by 29.1\% and 20.2\%.

We make three key contributions:
\begin{itemize}
    \item We propose a unified Physical--Digital execution model for the computing continuum that connects a running physical system with its continuously updated executable Digital Twin in the same runtime loop. Rather than treating the runtime and simulator as independent environments, Darpan allows digital exploration to begin directly from current physical execution and returns the selected operation to the real system for execution.
    \item We design synchronization and calibration driven by physical observations, immutable Snapshots that provide a common starting point for candidate comparison, isolated candidate execution, and feasibility validation against the latest physical state before execution. This allows multiple candidates to be compared under consistent conditions while preventing digital execution from feeding back into observations and models derived from the physical system.
    \item We implement and open-source Darpan and evaluate its prediction accuracy, adaptation, decision utility, recovery, scale-out, and runtime overhead using heterogeneous DAG workloads and a computing continuum with up to 40 nodes. An external DRL scheduling \casestudyref{} further demonstrates Darpan's ability to integrate external decision software and improve the original scheduler using Twin-generated experience under the same physical execution budget.
\end{itemize}

The rest of the paper is organized as follows. Section~\ref{sec:related} reviews continuum runtime frameworks, simulation and emulation tools, and Digital Twin approaches. Section~\ref{sec:design} presents Darpan's Physical--Digital execution model and closed runtime loop. Section~\ref{sec:implementation} describes the implementation and extension interfaces. Section~\ref{sec:evaluation} evaluates Darpan and reports the DRL \casestudyref{}. Section~\ref{sec:vision} presents future directions outlining the broader vision and research opportunities enabled by Darpan. Section~\ref{sec:conclusion} concludes the paper.

\section{Background and Related Work}
\label{sec:related}

Software for the computing continuum has developed along two lines~\cite{bittencourt2025continuum}. Runtime frameworks execute applications on real edge-to-cloud resources. Simulation and emulation tools evaluate policies in controlled virtual environments. In parallel, Digital Twin approaches have been studied for specific management tasks in the computing continuum. Each line addresses an important part of continuum operation, but none by itself makes the current state of a running application a reusable starting point for isolated alternatives and subsequent physical execution.

\subsection{Continuum Runtime Frameworks}

FogBus2 organizes application submission, scheduling, and container execution across heterogeneous IoT, edge, and cloud resources~\cite{deng2021fogbus2}. Its modular runtime discovers resources, profiles nodes and networks, and turns a selected placement into distributed physical execution. FogBus2 therefore establishes an end-to-end path from an application request to observable work on real resources. ReinFog extends that physical path with reusable Deep Reinforcement Learning (DRL) learners and workers~\cite{wang2025reinfog}. It collects node, network, and task information, exposes learning interfaces, and converts a learned scheduling decision into physical deployment. Its focus is the system integration needed to train and execute a resource-management policy. E2Clab standardizes experiments on edge-to-cloud testbeds~\cite{rosendo2020e2clab}. It covers configuration, application mapping, deployment, network control, measurement, and provenance, allowing resource-management policies to be repeated under controlled physical conditions. Unlike a runtime attached to one evolving application, its abstraction is an experiment whose configurations are prepared and replayed. These frameworks provide practical support for application deployment, execution, observation, and controlled experimentation on real edge-to-cloud resources. However, they primarily enact or evaluate selected decisions on the physical infrastructure and do not provide a reusable virtual execution point from the current state of a running application for exploring multiple isolated alternatives before physical execution.

\subsection{Continuum Simulation and Emulation Tools}

CloudSim provides a scalable discrete-event simulation framework for modeling datacenters, virtual machines, workloads, provisioning, and scheduling policies without deploying the corresponding physical infrastructure~\cite{calheiros2011cloudsim} with easily extensible simulation elements~\cite{clousim7g}. Building on CloudSim, iFogSim2 models multi-tier mobility, dynamic clustering, service migration, and microservice management~\cite{mahmud2022ifogsim2}. EdgeCloudSim models edge networking, device mobility, workload generation, and edge orchestration~\cite{sonmez2018edgecloudsim}, while PureEdgeSim covers heterogeneous cloud, edge, and mist resources together with mobility, networking, energy, and task orchestration~\cite{mechalikh2021pureedgesim}. ECLYPSE provides a unified Python interface for simulated, emulated, and hybrid cloud-edge experiments, where infrastructure and application graphs define the deployment and controlled events modify resource and network conditions~\cite{massa2026eclypse}. These tools provide controlled and repeatable environments for evaluating resource-management policies, but their virtual executions begin from experiment-defined infrastructure, workload, and policy configurations rather than from the current application progress, resource occupancy, artifact locations, and measurements of an application already running under a physical runtime.

\subsection{Digital Twin Approaches in the Computing Continuum}

Digital Twins have been increasingly applied to networked and distributed computing systems~\cite{qin2024dtnsurvey}. Existing approaches have studied self-healing~\cite{saxena2025selfhealing}, data access~\cite{gao2026cost}, and resource orchestration~\cite{li2026dtorchestration} in Digital Twin-enabled computing continua. These works demonstrate the value of keeping a digital representation aligned with a physical system. However, their Digital Twins are typically built around a particular management objective, and the digital representation serves that objective rather than acting as a general executable environment for a running continuum. They do not expose the current state of a running application as an immutable, reusable starting point for executing multiple isolated candidates, nor do they define how a selected result is revalidated against the latest physical state before physical execution. Because they are task-specific methods rather than general-purpose continuum frameworks, we do not include them in the framework-level comparison in Table~\ref{tab:comparison}. Darpan is complementary to them and can host such task-specific Digital Twin models and controllers through its extension interfaces.

\subsection{System Positioning}

Table~\ref{tab:comparison} compares general-purpose computing-continuum runtime and simulation frameworks against the capabilities needed for this execution model. Physical and virtual execution distinguish real work from controlled candidate execution. Online synchronization and calibration ask whether the virtual state and its performance models follow a running system. Snapshot branching asks whether competing candidates share one captured starting point without changing one another. Physical commit asks whether the selected result can be checked and applied through the physical runtime. Extensible integration and execution-plane scale-out cover external decision software and additional physical workers. Partial support denotes controlled real-code or hybrid experimentation without the complete live-state path.

\begin{table*}[!t]
\centering
\caption{Capability Comparison of Computing-Continuum Runtime and Simulation Frameworks}
\label{tab:comparison}
\vspace{2pt}
\footnotesize
\setlength{\tabcolsep}{1.5pt}
\renewcommand{\arraystretch}{1.12}
\resizebox{\linewidth}{!}{%
\begin{tabular}{@{}lcccccccc@{}}
\toprule
\textbf{Framework} & \makecell{\textbf{Physical}\\\textbf{Execution}} & \makecell{\textbf{Virtual}\\\textbf{Execution}} & \makecell{\textbf{Online}\\\textbf{Synchro-}\\\textbf{nization}} & \makecell{\textbf{Online}\\\textbf{Calibration}} & \makecell{\textbf{Snapshot}\\\textbf{Branching}} & \makecell{\textbf{Physical}\\\textbf{Commit}} & \makecell{\textbf{Extensible}\\\textbf{Integration}} & \makecell{\textbf{Execution-}\\\textbf{Plane}\\\textbf{Scale-Out}} \\
\midrule
FogBus2~\cite{deng2021fogbus2}       & \cmark & \xmark & \xmark & \xmark & \xmark & \xmark & \cmark & \cmark \\
ReinFog~\cite{wang2025reinfog}       & \cmark & \xmark & \xmark & \xmark & \xmark & \xmark & \cmark & \cmark \\
E2Clab~\cite{rosendo2020e2clab}      & \cmark & \pmark & \xmark & \xmark & \xmark & \xmark & \cmark & \cmark \\
CloudSim~\cite{calheiros2011cloudsim, clousim7g}& \xmark & \cmark & \xmark & \xmark & \xmark & \xmark & \cmark & \xmark \\
EdgeCloudSim~\cite{sonmez2018edgecloudsim} & \xmark & \cmark & \xmark & \xmark & \xmark & \xmark & \cmark & \xmark \\
iFogSim2~\cite{mahmud2022ifogsim2}   & \xmark & \cmark & \xmark & \xmark & \xmark & \xmark & \cmark & \xmark \\
PureEdgeSim~\cite{mechalikh2021pureedgesim} & \xmark & \cmark & \xmark & \xmark & \xmark & \xmark & \cmark & \xmark \\
ECLYPSE~\cite{massa2026eclypse}      & \pmark & \cmark & \xmark & \xmark & \xmark & \xmark & \cmark & \xmark \\
\textbf{Darpan (Proposed)}                       & \cmark & \cmark & \cmark & \cmark & \cmark & \cmark & \cmark & \cmark \\
\bottomrule
\end{tabular}
}
\vspace{2pt}
\footnotesize \cmark~= supported, \pmark~= partial support, \xmark~= not supported.
\end{table*}

Darpan maintains continuous state and execution semantics across observation of a live application, capture of one common state, isolated execution of alternatives, and a final physical feasibility check. It provides this complete process as a reusable runtime abstraction. Sections~\ref{sec:design} and~\ref{sec:implementation} describe its design and implementation, and Section~\ref{sec:evaluation} evaluates the resulting behavior.

\section{Darpan Runtime Design}
\label{sec:design}

\begin{figure*}[!t]
    \centering
    \includegraphics[width=\textwidth]{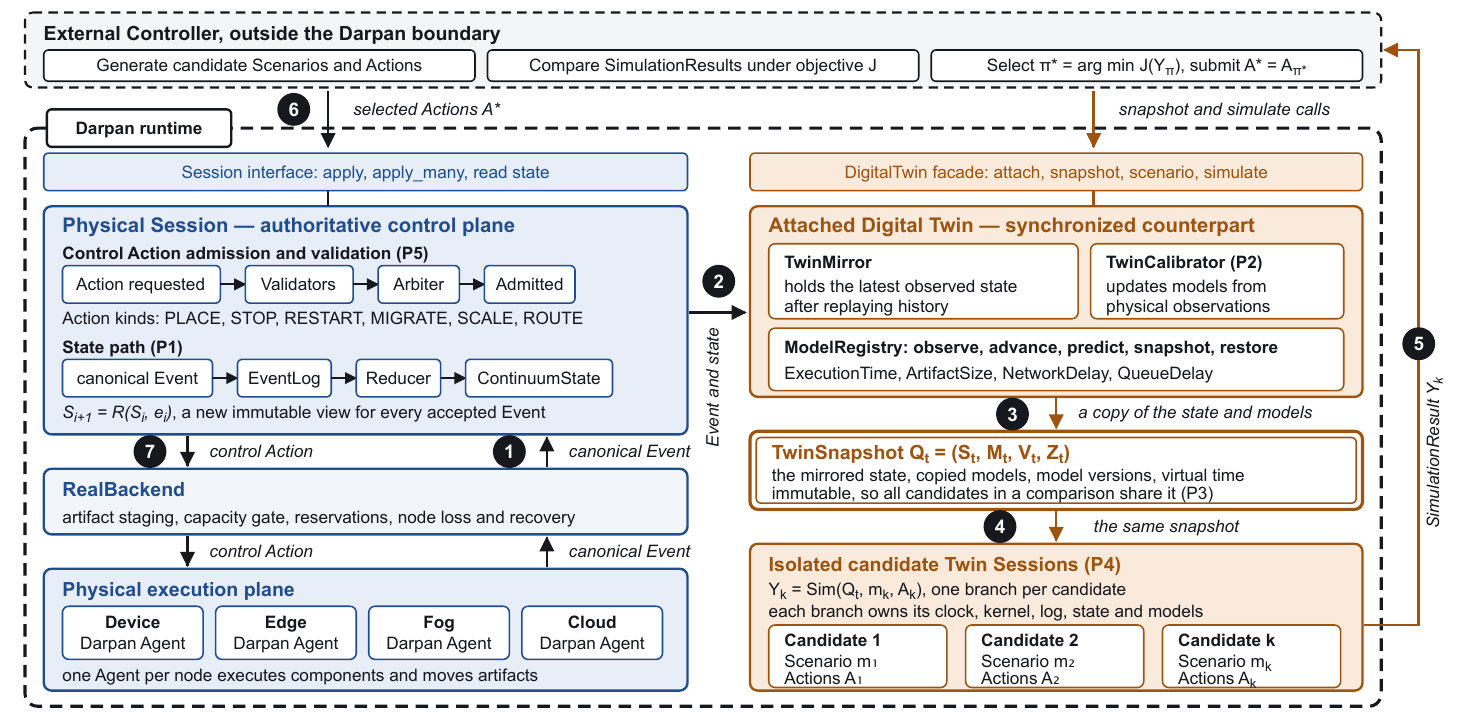}
    \caption{Darpan's closed physical and twin execution loop. The dashed frame encloses the Darpan runtime, which an external controller accesses through the Session interface and DigitalTwin facade. Physical execution produces canonical Events that update the authoritative ContinuumState~(1). TwinMirror synchronizes each Event together with its corresponding state, while TwinCalibrator updates the registered models using physical observations~(2). A TwinSnapshot captures the mirrored state, model states and versions, and virtual time~(3), providing the common immutable starting point for isolated candidate Twin Sessions~(4). Their SimulationResults return to the controller~(5), which submits only the selected Action sequence~(6). The Physical Session validates these Actions against the current ContinuumState before RealBackend dispatch~(7), closing the loop with new physical Events. Properties P1 through P5 are described in Section~\ref{subsec:properties}.}
    \label{fig:architecture}
\end{figure*}

Darpan maintains an attached executable Digital Twin for each running Physical Session. One Physical Session represents one managed continuum execution domain: it maintains the authoritative control-plane EventLog and ContinuumState for its participating nodes, resources, and concurrent applications. The attached Twin is its continuously synchronized predictive counterpart, while candidate Twin Sessions are temporary what-if branches created from snapshots. Physical computation, transfer, resource, and failure events are committed to the EventLog and reduced into the current ContinuumState. The attached Twin consumes the same event stream and calibrates its models from physical observations. An external decision program, called a Controller, may be a placement optimizer, recovery policy, or scheduler. The Controller captures a TwinSnapshot, executes candidate decisions in isolated Twin Sessions, compares their SimulationResults, and selects the operation returned to the physical system. Darpan represents each such operation as an Action, such as placing, migrating, or restarting a component. The Physical Session validates each Action against its authoritative state before RealBackend dispatch; the resulting physical Events begin the next synchronization cycle. Fig.~\ref{fig:architecture} summarizes this closed physical and twin execution path and the boundary between Darpan and external controllers.

\subsection{Physical and Twin Runtime Properties}
\label{subsec:properties}

Darpan separates two points in time. A TwinSnapshot defines a common starting point for comparing candidates, whereas the Physical Session provides the authoritative state for validating a selected Action. Physical applications and resources may continue to evolve while candidate Sessions execute; a snapshot therefore does not lock the physical system. Darpan preserves five runtime properties throughout this Physical and Twin loop.

\emph{P1: Authoritative state reconstruction.} Facts accepted by a Session are represented as canonical Events and appended to its EventLog. The Session's authoritative ContinuumState evolves only through the Reducer, so replaying the same ordered Event sequence reconstructs the same application progress, node status, artifact locations, and resource occupancy. Counterfactual Scenario transformations instead derive a private candidate starting state from a Snapshot; they modify neither the source Snapshot nor the Physical Session.

\emph{P2: Physical-only calibration.} Only valid observations emitted by the Physical Session update the models of the attached Twin. Virtual completions and transfers produced by candidate Twin Sessions remain in their own SimulationResults. They never feed back into the attached Twin, preventing a model from recursively learning from its own predictions.

\emph{P3: Immutable fork configuration.} A TwinSnapshot copies the latest mirrored ContinuumState together with the serialized model states, recorded version identifiers, and virtual time available at capture. Every candidate being compared is created from this same captured state and model configuration, so result differences arise from the candidate conditions and Actions rather than from different application progress or model snapshots.

\emph{P4: Isolated candidate execution.} Each candidate owns a Session, StateStore, EventLog, ModelRegistry, VirtualClock, and SimulationKernel. Its events, resource changes, and model evolution cannot modify the source Snapshot or become visible to another candidate. Candidates are logically independent branches, not sequential modifications of a shared simulator state.

\emph{P5: Submission-time feasibility validation.} A Twin result never replaces physical state. The controller returns a canonical Action, which the Physical Session checks against the authoritative ContinuumState that exists at submission time. This validation establishes execution feasibility under the current lifecycle, resource, reachability, and backend constraints; objective definition and candidate selection remain Controller responsibilities. The Controller can respond to a rejected or failed Action by initiating a new snapshot and selection round. Accepted Actions remain observable through canonical started, completed, or failed Events as RealBackend performs the physical operation.

Together, these properties give the Snapshot a fixed basis for candidate comparison and the Physical Session authority over current-state validation and physical execution. New physical Events do not retroactively change a Snapshot, but they update the attached Twin and become the basis for subsequent Snapshots.

\subsection{Unified Execution Model}

We represent a computing continuum as
\begin{equation}
    \mathcal{C}=(\mathcal{N},\mathcal{L}),
    \label{eq:continuum}
\end{equation}
where $\mathcal{N}$ is a set of heterogeneous compute nodes and $\mathcal{L}$ is a set of network links. Node state describes processing capacity, memory, storage, availability, resource occupancy, and measurements. Link state describes endpoints, reachability, latency, bandwidth, and measurements. Named resources allow accelerators and other capacities to be added without changing the core state representation.

An application is a Directed Acyclic Graph (DAG)
\begin{equation}
    G=(V,E),
    \label{eq:dag}
\end{equation}
where $V$ contains independently deployable tasks or services and $E$ defines execution dependencies and artifact transfers. A placement $\pi:V\rightarrow\mathcal{N}$ maps components to nodes subject to resource, lifecycle, and network constraints. A Workload is a timed sequence of application arrivals, allowing physical and twin runtimes to execute the same application and arrival descriptions.

A \emph{Session} is Darpan's execution boundary. One Physical Session covers a managed continuum domain and records the control-plane facts and measurements of its concurrent applications and participating resources in one authoritative EventLog and ContinuumState. A Twin Session records the virtual Events and final state of one candidate branch. Each Session owns a Clock, EventLog, StateStore, validators, arbiter, observers, and one RuntimeBackend. Physical and Twin Sessions use WallClock and VirtualClock, respectively, while sharing SystemSpec, ApplicationSpec, WorkloadSpec, Event, ContinuumState, and Action semantics.

Let $e_i$ denote the $i$th canonical Event and $R$ the Reducer. Authoritative state within a Session evolves only as
\begin{equation}
    S_{i+1}=R(S_i,e_i).
    \label{eq:state-transition}
\end{equation}
ContinuumState is immutable: the Reducer creates a new materialized view for every accepted Event. Events carry event and receive times, source sequence numbers, and correlation and causation identifiers. The EventLog therefore preserves traceable evidence, while ContinuumState provides an efficient current view for controllers and backends.

A Scenario transformation has a different role: it derives a private candidate starting state from a copied Snapshot before a candidate Session begins. Once instantiated, the candidate Session evolves that private state through its own canonical Events and Reducer. The authoritative Physical Session and the source Snapshot remain unchanged.

An Action expresses control intent rather than an already completed state change. A controller proposes the Action, the Session validates and arbitrates it, the backend executes it, and the outcome re-enters the Session as canonical Events. Darpan's closed execution contract comprises two coupled paths:
\begin{equation}
\begin{aligned}
\mathit{Controller} &\rightarrow \mathit{Action} \rightarrow \mathit{Session}
\rightarrow \mathit{Backend},\\
\mathit{Backend} &\rightarrow \mathit{Event} \rightarrow \mathit{EventLog}\\
&\rightarrow \mathit{Reducer} \rightarrow \mathit{ContinuumState}.
\end{aligned}
\label{eq:execution-contract}
\end{equation}
ContinuumState is read through the Session boundary by backends, TwinMirror, and external controllers, thereby closing the loop. Because physical and virtual executions implement the same canonical Event, state, Action, and Session semantics, a controller can use either path without translating between unrelated simulator and runtime representations. Darpan does not prescribe candidate generation or an objective. Placement, recovery, scheduling, and learning controllers derive objectives supported by registered models, while additional objectives can be incorporated through model extensions.

\subsection{Physical Execution and State Reconstruction}

The Physical Runtime combines a Physical Session, RealBackend, and Darpan Agents deployed across Device, Edge, Fog, and Cloud nodes. An Action first enters the Session's common validation path. Validators check whether the active backend supports the Action, whether the target node has a configured executor, whether the component lifecycle permits the request, whether resources are feasible, and whether predecessor artifacts can reach the target. When simultaneous Actions conflict on the same target, an arbiter admits the highest-priority compatible request.

RealBackend maps an admitted component Action to asynchronous execution at the target Agent. The Agent reserves local resources, starts a process or container, supports waiting and cancellation, and reports execution status, duration, artifacts, and resource observations. For cross-node dependencies, RealBackend stages required input artifacts before allocating compute resources. A DAG successor becomes ready only after all predecessors have completed and its required artifacts are reachable.

Physical Event commitment is serialized. The backend may first perform a physical preparation step; the Session then appends the Event idempotently to the EventLog, invokes the Reducer through StateStore, records the associated materialized state, and finally notifies TwinMirror and other observers. The DAG orchestrator emits causally linked component-ready, dependency-release, and application-completion Events through the same path. Thus, every observer sees a state that already includes the current Event, while a retried physical report cannot be applied twice.

Task starts and completions, resource allocation and release, artifact transfers, and node and link measurements all become canonical Events. Completion Events preserve node, execution interval, and resource evidence; transfer Events preserve endpoints, route, artifact size, and physical duration. These Events simultaneously reconstruct physical state and supply synchronization and calibration evidence to the attached Twin.

When sustained connectivity loss makes an Agent unavailable, ClusterMonitor records the node as offline. RealBackend prevents new dispatch to that node, cancels the corresponding controller-side operations and transfers, releases backend reservations and canonical allocations, and marks affected work failed in the Session's control-plane state. Loss of the control connection is not treated as evidence that an already-started remote process has terminated. Node recovery restores availability but does not silently replay failed work. Retry or recovery requests re-enter the normal readiness, validation, and execution path, leaving an auditable fault history instead of hiding recovery inside the backend.

\subsection{Twin Synchronization and Online Calibration}

A Digital Twin attaches to a specific Physical Session. At attachment, TwinMirror can replay the Session history to recover nodes, links, application progress, active components, and artifact locations. It then subscribes to each new $(\mathit{event},\mathit{state})$ pair and retains the latest physical ContinuumState. The Physical Session remains the source of truth; TwinMirror maintains its most recently observed view.

TwinMirror also forwards the physical Event and its post-event state to TwinCalibrator. ModelRegistry manages ExecutionTime, ArtifactSize, NetworkDelay, and QueueDelay models under a common lifecycle with \texttt{observe}, \texttt{advance}, \texttt{predict}, \texttt{snapshot}, and \texttt{restore} operations. Each model consumes only relevant physical Events. Samples marked as contention-affected may be retained for end-to-end analysis but are not written again into a single-operation baseline, which avoids counting the same contention effect in both model prediction and simulated scheduling.

For a valid physical observation $x_t$, the execution model updates its mean and variance as
\begin{align}
    \mu_t &= \mu_{t-1}+\alpha(x_t-\mu_{t-1}),
    \label{eq:mean-update}\\
    \sigma_t^2 &= (1-\alpha)\left[\sigma_{t-1}^2+
    \alpha(x_t-\mu_{t-1})^2\right],
    \label{eq:var-update}
\end{align}
where $\alpha$ controls the contribution of new evidence. All per-key exponential updates evaluated in this paper use $\alpha=0.25$. The ArtifactSize model learns observed component outputs, while the QueueDelay model separates waiting for resources from baseline execution time.

The network model combines the current topology, link measurements, and completed physical transfers. For an artifact of size $b$ traversing path $p$, base transfer time is
\begin{equation}
    d(p,b)=\sum_{l\in p}\left(\delta_l+\frac{8b}{B_l}\right),
    \label{eq:network}
\end{equation}
where $\delta_l$ and $B_l$ are the latency and effective bandwidth of link $l$. Canonical link measurements override the corresponding configured parameters, and completed transfers calibrate residual endpoint or path error. Artifact-size and network uncertainty remain available to Twin execution rather than collapsing to constants after the first observation.

A prediction uses only observations committed before that prediction request. When the corresponding physical execution later completes, its result updates only subsequent model states; already returned predictions and created Snapshots do not change. Events generated by Twin Sessions do not pass through TwinMirror. This ordering makes comparisons between physical and twin execution predict-then-observe executions and ensures that adaptation reflects new physical evidence rather than self-generated virtual data.

\subsection{Versioned Snapshot and Isolated Scenario Execution}

A TwinSnapshot is an immutable candidate starting point
\begin{equation}
    Q_t=(S_t,M_t,V_t,Z_t),
    \label{eq:snapshot}
\end{equation}
where $S_t$ is the latest ContinuumState observed by TwinMirror when capture begins, $M_t$ the model states copied during capture, $V_t$ their recorded version identifiers, and $Z_t$ virtual time. The resulting tuple is immutable and fixes the captured state and model input used by every candidate in a comparison.

The controller constructs candidates from a common $Q_t$. A Scenario contains a source Snapshot, ordered counterfactual transformations, injected Events, a seed field, and an optional horizon. Current transformations include node removal, link changes, and physical-value changes. They are applied only to a copy of $S_t$ to form the candidate's private initial state. Placement, migration, scaling, and routing choices are expressed as Actions applied to the resulting Twin Session, so candidate decisions still traverse the common Session path. Scenario transformations define private what-if conditions and are never submitted to the Physical Session.

For candidate modification $m_k$ and Action sequence $A_k$, Darpan computes
\begin{equation}
    Y_k=\operatorname{Sim}(Q_t,m_k,A_k).
    \label{eq:scenario}
\end{equation}
Session creation copies ModelRegistry and restores $M_t$, creates a VirtualClock at $Z_t$ and a fresh SimulationKernel, and initializes an independent StateStore and EventLog from the materialized Scenario state. TwinBackend then executes workload arrivals, Actions, DAG dependencies, component computation, resource waiting, concurrent artifact transfers, release, and failure Events. The resulting SimulationResult contains the candidate's final ContinuumState and canonical Event sequence.

The source snapshot remains unchanged. Events from candidate $k$ cannot change the state, models, or clock of candidate $j$, and cannot enter the Physical EventLog. A controller may execute candidate Sessions sequentially or concurrently; semantically, all begin from the same $Q_t$. Placement, recovery, and scheduling decisions can consequently be compared without reconstructing application progress and resource state separately for each candidate.

The Physical Session may continue to receive completions, resource changes, and failures during candidate execution. These Events update the attached Twin but do not retroactively modify $Q_t$. The snapshot provides consistency among candidates, not a lock on physical execution. Darpan handles the gap between snapshot state and current physical state through the return protocol below.

\subsection{Validated Action Return and Controller Interface}

Given candidate set $\Pi$, SimulationResult $Y_\pi$, Action sequence $A_\pi$, and objective $J$, an external controller may select
\begin{equation}
\begin{aligned}
    \pi^*&=\arg\min_{\pi\in\Pi} J(Y_\pi),\\
    A^*&=A_{\pi^*}.
\end{aligned}
    \label{eq:selection}
\end{equation}
Outcome comparison is not an internal Twin module. A controller may use rules, search, mathematical optimization, or single- or multi-objective learning, and may evaluate only a small domain-informed candidate set. SimulationResults expose final state and canonical Events from which response time, resource waiting, and transfer time can be derived. Controllers may consume additional metrics when the corresponding models are registered.

The selected result does not overwrite the Physical Session. The controller submits only $A^*$ through \texttt{apply} or \texttt{apply\_many}; counterfactual Scenario transformations remain inside the candidate branch. The Session records \texttt{ACTION\_REQUESTED}, performs capability, node-mapping, lifecycle, resource-feasibility, and reachability checks on the current physical state, and records \texttt{ACTION\_ACCEPTED} or \texttt{ACTION\_REJECTED}. The arbiter removes conflicts among accepted Actions. Selected Actions are dispatched in order, and each dispatched request receives its own \texttt{ACTION\_STARTED} and \texttt{ACTION\_COMPLETED} or \texttt{ACTION\_FAILED} Events. Dependent control chains can be staged as the preceding Action produces new Events and state.

If a Snapshot is captured at $t$ and an Action is submitted at $t'>t$, its submission condition is
\begin{equation}
    \operatorname{Valid}(a,S_{t'})=\mathrm{true},
    \label{eq:latest-validation}
\end{equation}
not validity against $S_t$. We call $\Delta=t'-t$ the snapshot-to-submit interval. A node that went offline, resources acquired by another application, an advanced component lifecycle, or a broken data path can therefore invalidate an otherwise useful Twin result at submission. The check establishes current execution feasibility; the Controller remains responsible for the objective and selection policy. Rejection or execution failure may trigger a new snapshot and candidate-selection round. After acceptance, RealBackend reads the current physical state for the operation, reserves placement capacity where applicable, and reports completion or failure through canonical Events. These new physical facts close the execution, synchronization, calibration, branching, and return loop. \expref{7} measures how often a decision becomes infeasible as $\Delta$ and the physical change rate increase, and the time required to check and submit a predefined backup after rejection.

Together, these mechanisms form Darpan's closed physical and Twin loop. Physical execution produces observations that update the Twin; a Snapshot supplies the same starting point to isolated candidate executions; the external Controller compares their results; and the selected Action is checked against the current physical state before execution creates the next observations. This loop realizes the Physical Execution, Virtual Execution, Online Synchronization, Online Calibration, Snapshot Branching, and Physical Commit capabilities summarized in Table~\ref{tab:comparison}.

\section{Implementation}
\label{sec:implementation}

\begin{figure*}[!t]
    \centering
    \includegraphics[width=0.8\textwidth]{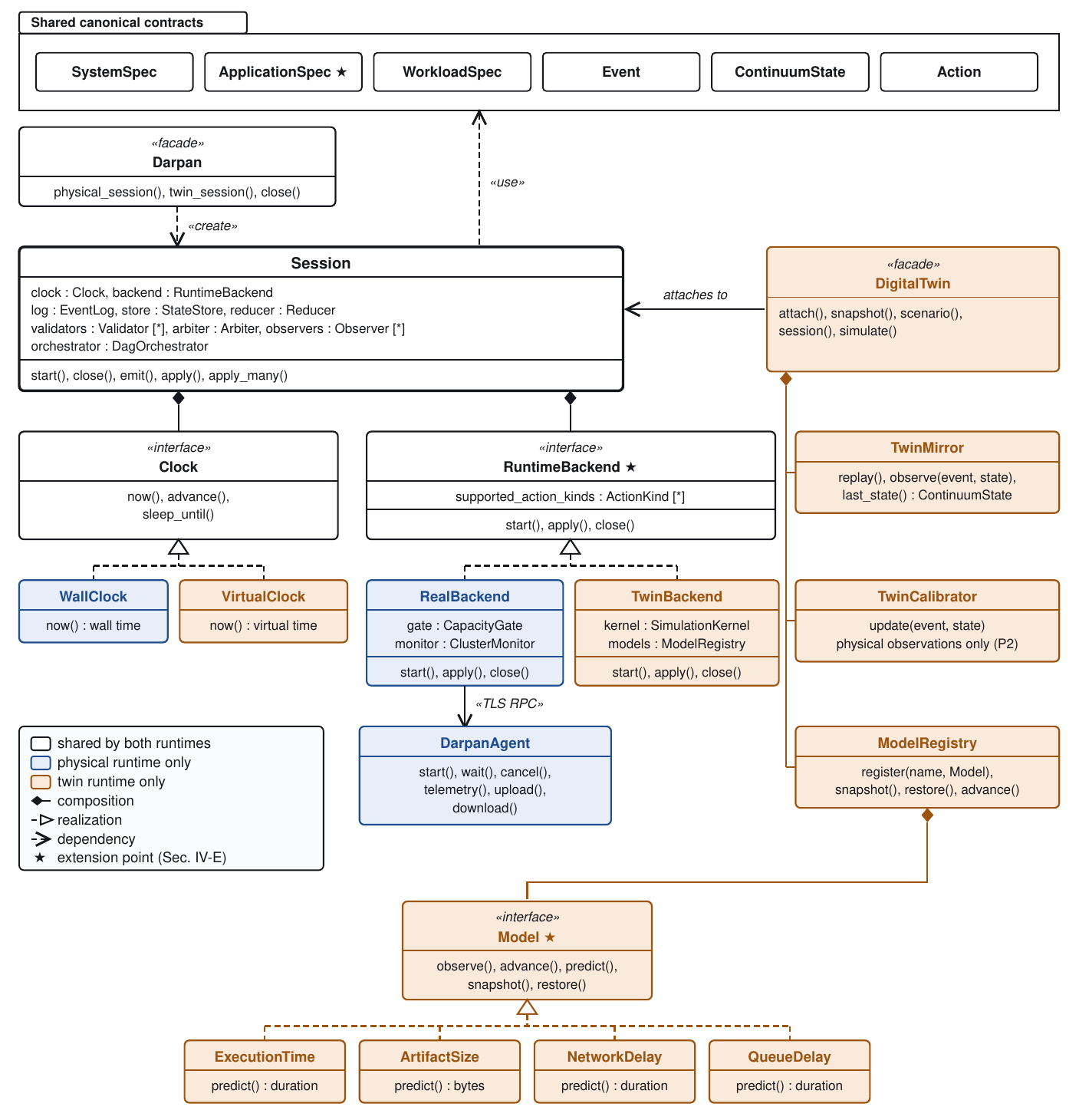}
    \caption{Implementation structure of Darpan. A shared Session composes a Clock and RuntimeBackend, with Physical and Twin realizations for each. DigitalTwin attaches to a Physical Session and composes TwinMirror, TwinCalibrator, and ModelRegistry. Shared canonical contracts and extension interfaces allow physical and virtual execution to use the same application, state, Event, and Action semantics.}
    \label{fig:implementation}
\end{figure*}

Darpan Core is implemented in Python. Rather than maintaining separate physical-runtime and simulator control stacks, the implementation realizes the contract in Eq.~\eqref{eq:execution-contract} with a small set of shared objects. RealBackend and TwinBackend use the same Session, application, Event, state, and Action interfaces; all state changes pass through canonical Events and the Reducer. Fig.~\ref{fig:implementation} summarizes this implementation structure, including the shared interfaces, physical and Twin runtime realizations, and extension points.

\subsection{Core Runtime and Public Interface}

The public \texttt{Darpan} facade creates and closes Physical and Twin Sessions. A Physical Session combines WallClock and RealBackend; a Twin Session combines VirtualClock and TwinBackend. The \texttt{DigitalTwin} facade exposes \texttt{attach}, \texttt{snapshot}, \texttt{scenario}, \texttt{session}, and \texttt{simulate}, allowing an external controller to observe a Physical Session, capture current state, construct a candidate, and obtain its SimulationResult. Candidate generation and outcome selection remain outside this facade.

Session is the core execution object. It owns a backend, Clock, EventLog, StateStore, application orchestrator, validators, arbiter, observers, and managed background tasks. \texttt{Session.start} supplies the backend with \texttt{emit}, \texttt{state}, and \texttt{now} callbacks; hence, a backend reads current state and reports outcomes only through the Session boundary. \texttt{Session.close} shuts down services and the backend in reverse order, cancels outstanding managed tasks, and releases observer references so that a completed candidate cannot affect later execution.

\texttt{Session.emit} implements the Event-commit order. A backend may first perform preparation needed to make a physical change effective. An emit lock then serializes EventLog append, Reducer application through StateStore, history indexing, and condition notification. In-memory and newline-delimited JavaScript Object Notation (JSON) EventLogs both reject duplicate Event identifiers; the persistent implementation supports replay after restart. Observers receive the materialized post-event state outside the lock, preventing calibration, monitoring, or controller callbacks from blocking canonical state commitment. Causal child Events emitted by the application orchestrator re-enter the same path.

\texttt{Session.apply\_many} implements a batch of per-Action lifecycles. Every request first emits \texttt{ACTION\_REQUESTED}. Validators return acceptance or a rejection reason against the same latest state; a priority arbiter retains a compatible set for each target. Selected Actions are then dispatched in order, and every backend invocation is enclosed by its own \texttt{ACTION\_STARTED} and completion or failure Events. The method is not a transactional multi-Action commit: an execution failure stops the remaining batch without rolling back Actions already completed. Dependent chains can instead be submitted in stages as new Events materialize updated state. Physical and Twin runtimes share this Action protocol.

Darpan's stable Action vocabulary comprises \texttt{PLACE}, \texttt{STOP}, \texttt{RESTART}, \texttt{MIGRATE}, \texttt{SCALE}, and \texttt{ROUTE}. An ActionKind expresses canonical intent but does not imply support by every backend. Each backend advertises \texttt{supported\_action\_kinds}; the capability validator rejects unsupported Actions before dispatch.

\subsection{Physical Backend and Darpan Agents}

RealBackend maps admitted Actions to local executors or authenticated remote Darpan Agents. The physical cluster uses a strict node-to-Agent mapping: a node without a configured executor is rejected rather than silently executed on the controller host. The Agent control plane supports authenticated and Transport Layer Security-protected asynchronous start, wait, cancellation, telemetry, artifact upload and download, and optional inter-Agent measurement and direct artifact forwarding.

A component \texttt{PLACE} resolves its ApplicationSpec and target, stages input artifacts, and then enters an asynchronous resource-capacity gate. The gate combines declared SystemSpec capacity, the latest physical capacity measurement, and temporary RealBackend reservations not yet visible in canonical state. It therefore prevents concurrent Actions that validated against the same state from jointly oversubscribing a node. After admission, RealBackend emits resource-allocation and component-start Events and delegates the process or container to the target executor. Completion collects exit status, duration, and produced artifacts, releases capacity, and emits a component-completed or component-failed Event.

Cross-node file dependencies cause real data movement before computation. By default, the controller relays bounded chunks between source and target Agents. When direct forwarding is explicitly enabled, the target issues a short-lived upload ticket bound to path and size; the source does not receive long-term target credentials. Every transfer produces started, completed, or failed Events with endpoints, route, size, physical duration, and contention metadata. These Events release DAG dependencies and calibrate the network model.

ClusterMonitor converts sustained Agent connectivity changes into \texttt{NODE\_OFFLINE} and \texttt{NODE\_RECOVERED} Events. On loss of connectivity, RealBackend prevents new dispatch, cancels controller-side execution handles and transfers involving that node, releases backend reservations and canonical allocations, and records failed work in the Session's control-plane state. Because the remote Agent is unreachable, Darpan does not infer that an already-started remote process has terminated. Recovery restores availability only. Bounded retry for finite DAG tasks follows ApplicationSpec; service restart and migration remain explicit Actions. Both routes pass through Session validation instead of concealing recovery inside RealBackend.

RealBackend also permits a constrained PhysicalControlDriver to apply approved CPU or link changes before their corresponding Events are committed. The Event is recorded only if the driver confirms that the physical modification succeeded, preserving the EventLog as a record of physical facts.

\subsection{Twin Execution Engine and Model Lifecycle}

TwinBackend implements the same RuntimeBackend protocol and the six canonical Action kinds. It uses a VirtualClock and replaceable SimulationKernel to process time-ordered Events. Actions, workload arrivals, and Scenario-injected Events enter the same discrete-event queue. The clock advances only when the next Event is removed, and ModelRegistry advances over the same virtual interval. Twin execution therefore avoids wall-clock waiting while preserving deterministic event order.

TwinBackend composes DAG execution, artifact transfer, and resource use. ExecutionTime supplies component baselines; ArtifactSize supplies transfer sizes; NetworkDelay supplies path costs under current link measurements; QueueDelay represents waiting. SimulationKernel adds DAG dependencies, strict resource reservation, optional CPU fair sharing, concurrent link transfers, and failure cancellation. When computation or transfer membership changes, or a link measurement changes, the compute and network schedulers recompute outstanding completion times instead of assuming independent, dedicated resources.

TwinBackend emits physical-semantic component, resource, transfer, Action, and failure Events. If a node or link fails, queued completion Events are invalidated and replaced with failure Events; recovery does not reactivate failed work. SimulationResult consequently preserves an auditable execution trace as well as the final state.

ModelRegistry gives each model the common \texttt{observe}, \texttt{advance}, \texttt{predict}, \texttt{snapshot}, and \texttt{restore} lifecycle. New models register by name without changes to TwinMirror, TwinSnapshot, or Session. The Registry of an attached Twin and those of candidate Sessions are distinct objects. A candidate restores parameters from a snapshot; its virtual observations and time advancement cannot mutate the attached Registry.

\subsection{Snapshot and Branch Isolation}

\texttt{DigitalTwin.attach} uses Session identity to prevent duplicate attachment. With history replay enabled, each recorded $(\mathit{event},\mathit{state})$ pair is delivered to TwinMirror in order before a live observer is registered. \texttt{DigitalTwin.snapshot} reads \texttt{TwinMirror.last\_state}, or an explicitly supplied state, and copies the ModelRegistry state available during the call together with virtual time, recorded model-version identifiers, and the snapshot-schema version. The captured state and model combination is then fixed for all candidates that use the Snapshot.

A Scenario stores its source TwinSnapshot, ordered modifications, injected Events, seed, and horizon, and supports JSON serialization. \texttt{Scenario.materialize} applies modifications to a copy of Snapshot state and returns the candidate's initial ContinuumState without mutating the snapshot. Candidate Actions are then applied through the newly created Session and its normal Action path.

\texttt{DigitalTwin.session} is the concrete isolation boundary. It copies the attached ModelRegistry, restores \texttt{model\_states} from the snapshot, constructs a new VirtualClock, SimulationKernel, and TwinBackend, and creates independent StateStore and EventLog objects from the materialized Scenario. Candidate Sessions share no mutable execution object. Closing one candidate clears its backend queue, schedulers, cancellation state, and managed tasks without affecting the snapshot, attached Twin, or another candidate.

SimulationResult contains only the candidate's final ContinuumState and canonical Event sequence. A controller can derive response time, transfer overhead, and resource waiting from those Events or inspect final constraints directly. Snapshot and Scenario are persisted alongside SimulationResult as candidate provenance, preserving the captured fork point, model versions, and counterfactual modifications used for that execution.

\subsection{Extension Interfaces}

Applications extend Darpan through ApplicationSpec, WorkloadSpec, and component executors. A DAG configuration and its physical process or container are sufficient; no application-specific class is added to Core. Physical and Twin runtimes read the same ApplicationSpec, while RealBackend executes components and transfers and TwinBackend executes the corresponding modeled semantics.

A backend extension implements \texttt{start}, \texttt{apply}, and \texttt{close} and declares supported Action kinds. A new stable Action requires a schema, validator, canonical Events, Reducer semantics, physical and twin implementations, and contract tests. This requirement prevents an enum value without comparable physical and virtual behavior from appearing as a supported feature.

A model extension implements the Registry lifecycle above. A controller depends only on current state, TwinSnapshot, Scenario, SimulationResult, and Action interfaces. A Reinforcement Learning adapter defines observation mapping, Action conversion, reward, Action space, and which ready decisions are learning-controlled; ordinary policies can handle all other decisions. Darpan Core does not depend on a particular DRL library, and all scheduling methods in the \casestudyref{} remain external controllers.

\subsection{Execution-Plane Scale-Out and Reproducibility}

The current scale-out boundary is the physical execution plane. One authoritative Physical Session control plane coordinates applications, Agents, and transfers through asynchronous tasks, while components execute independently at their nodes. Session serializes only canonical Event commitment and state materialization; it does not hold a global lock during physical computation or network transfer. A node joins by registering Agent identity, address, resources, and capabilities in ClusterInventory; ApplicationSpec and controller interfaces remain unchanged. RealBackend maintains resource reservations per node, and each Session owns its EventLog and StateStore.

Each run persists resolved SystemSpec, ApplicationSpec, WorkloadSpec, and Scenario inputs, random seeds, the canonical EventLog, final state, model snapshot, and software and environment provenance. Input and local-plugin checksums are recorded; physical runs preserve Agent protocol and environment fingerprints, while candidate records preserve their source snapshot and model versions. Matched experiments can therefore execute with the same frozen inputs and seeds while retaining both raw observations and aggregates.

These implementation paths provide the Extensible Integration and Execution-Plane Scale-Out capabilities in Table~\ref{tab:comparison} while realizing the runtime properties of Section~\ref{sec:design} without evaluation-specific execution engines.

\section{Performance Evaluation}
\label{sec:evaluation}

We organize the evaluation around seven questions. \expref{1} asks whether the Twin follows physical execution, and \expref{2} whether it adapts when resources and networks change. \expref{3} tests whether candidate execution improves placement, while \expref{4} follows the complete loop during failure recovery. \expref{5} measures physical execution as the cluster grows, and \expref{6} measures the time required to maintain and use the Twin. \expref{7} studies decisions that become invalid because the physical state changes between selection and execution. The \casestudyref{} then asks whether Darpan can incorporate an external scheduler and whether its generated experience improves that scheduler under the same physical-DAG budget.

\subsection{Experimental Setup}

\subsubsection{Physical Continuum}
\begin{figure*}[!t]
    \centering
    \includegraphics[width=0.96\textwidth]{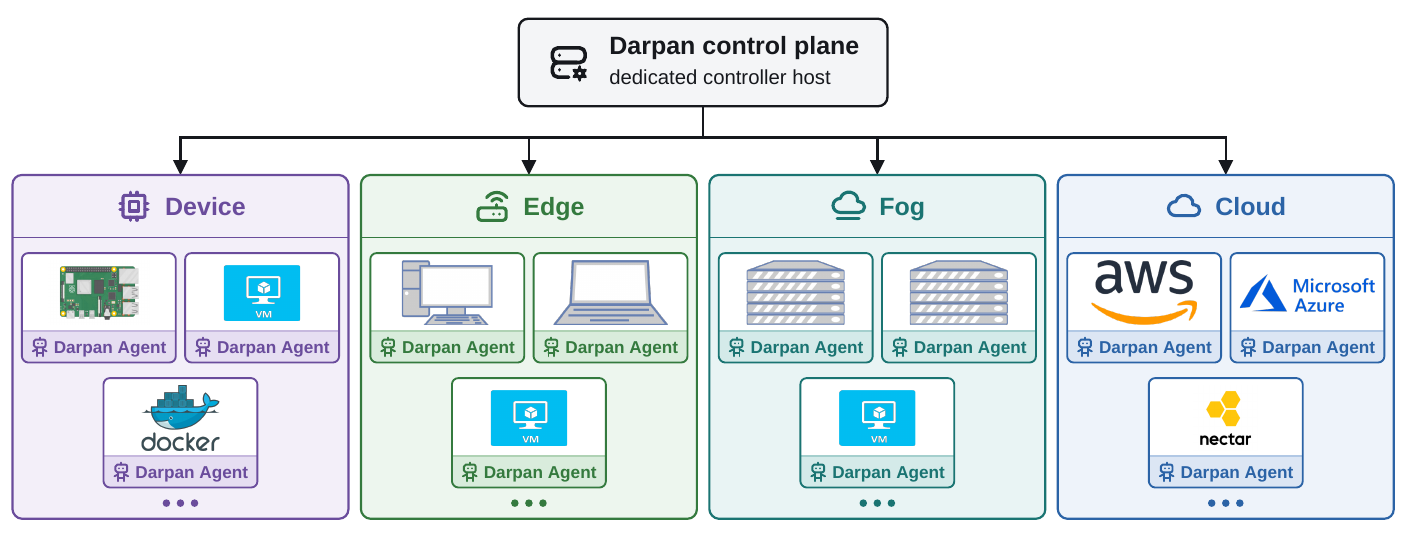}
    \caption{Deployment of Darpan on the physical continuum. The Darpan control plane runs on a dedicated controller host, and every Device, Edge, Fog, and Cloud node runs a Darpan Agent.}
    \label{fig:deployment}
\end{figure*}

To evaluate Darpan's support for heterogeneous computing continua, we construct a multi-tier environment comprising Device, Edge, Fog, and Cloud resources, as illustrated in Fig.~\ref{fig:deployment}. The Device tier includes Raspberry Pi devices (quad-core Broadcom BCM2837 at 1.2~GHz with 1~GB of memory), virtual machines, and Docker containers. The Edge tier includes systems based on Intel Core i7 processors (6 cores at 3.2~GHz with 8~GB of memory) and Apple M1 processors (8 cores with 16~GB of memory), together with two AMD EPYC virtual machines (2~vCPUs at 2.0~GHz with 8~GB of memory). The Fog tier includes systems based on Intel Core i7 processors (14 cores at 2.3~GHz with 16~GB of memory) and Intel Core i9 processors (8 cores at 2.5~GHz with 32~GB of memory), together with two AMD EPYC virtual machines (4~vCPUs at 2.0~GHz with 16~GB of memory). The Cloud tier spans Amazon Web Services, Microsoft Azure, and Nectar, using AMD EPYC or Intel Xeon instances ranging from 2~vCPUs at 2.0~GHz with 8~GB of memory to 32~vCPUs at 3.5~GHz with 128~GB of memory. Inter-node link capacities range from 100~Mbit/s to 40~Gbit/s. Every physical or virtual node runs a Darpan Agent that executes DAG components, transfers artifacts, collects compute and network observations, and applies resource and failure controls.

\subsubsection{DAG Workloads}
We use six DAG workloads. \emph{CPU Analytics} models a video-analysis pipeline with preprocessing, parallel feature extraction, fusion, and inference, and is used in \expref{1} and \expref{4}. \emph{Data Transform} models splitting, compression, checksumming, merging, and publishing, and is used in \expref{1}. \emph{Stream Window} models parallel filtering and feature analysis followed by detection aggregation, and is used in \expref{1} and \expref{3}. \emph{Drift Probe} is sensitive to both node service rate and cross-tier transfer performance and is used in \expref{2}. \emph{Scale DAG} is a fixed serial compute pipeline whose concurrency grows with node count in \expref{5}. \emph{Wide Parallel Scheduling} combines broad parallelism with multi-stage aggregation and is used to time Twin candidate execution in \expref{6}, to measure the effect of evaluating more candidates in \expref{7}, and in the DRL \casestudyref{}. \expref{7} additionally uses a small placement probe to measure the time required to reject an infeasible placement and submit a predefined backup.

\subsection{E1: Physical and Twin Fidelity}
\label{sec:e1}

\expref{1} evaluates whether the Physical Runtime and the attached Twin produce consistent end-to-end behavior. We execute a random continuous sequence of 100 CPU Analytics, Data Transform, and Stream Window DAGs. Before every physical execution, the Twin predicts response time using the same DAG, placement, and model state available at that point.

\begin{figure*}[!t]
    \centering
    \includegraphics[width=0.96\textwidth]{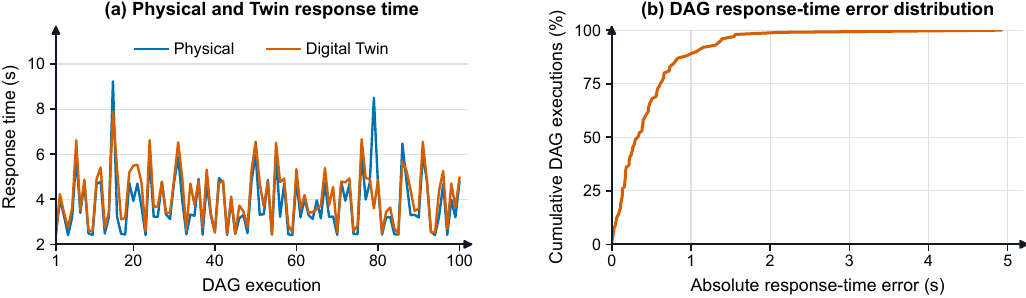}
    \caption{Fidelity between physical and twin execution over 100 continuous DAG executions. (a) Physical and twin response times for the same DAG and placement. (b) Absolute-error distribution over the same executions.}
    \label{fig:e1}
\end{figure*}

Fig.~\ref{fig:e1}(a) shows that the Physical and Twin curves follow the same changes across workloads and placements. Their response times have a correlation coefficient of 0.849 and a mean absolute error of 0.485~s. Fig.~\ref{fig:e1}(b) shows that 90\% of absolute errors are at most 1.074~s and 95\% are at most 1.375~s. The executable twin therefore preserves the dominant variation of physical DAG execution and provides a useful basis for comparing candidate scenarios.

\subsection{E2: Online Adaptation}
\label{sec:e2}

\expref{2} tests whether synchronization and calibration follow physical changes that are not announced to prediction methods. Drift Probe passes through Normal, CPU Pressure, Network Pressure, Mixed Pressure, and Recovery phases. CPU Pressure reduces available Cloud compute service; Network Pressure degrades transfer performance between Fog and Cloud; Mixed Pressure applies both changes. Each phase executes 16 DAGs, and the complete process is independently repeated ten times.

\begin{figure*}[!t]
    \centering
    \includegraphics[width=0.96\textwidth]{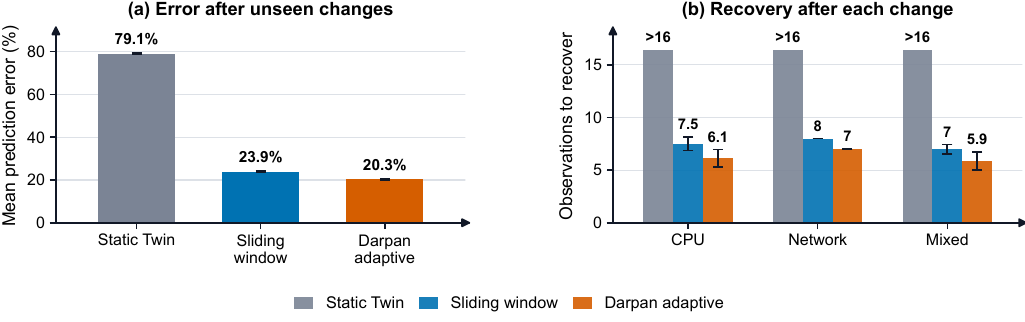}
    \caption{Online adaptation after unannounced CPU, network, and mixed changes. (a) Mean prediction error over the adaptation periods. (b) Physical observations required to reach the recovery criterion.}
    \label{fig:e2}
\end{figure*}

We compare Static Twin, Sliding Window, and Darpan Adaptive Twin. Static Twin retains its initial model. Sliding Window estimates current performance from the eight most recent physical executions. Darpan continuously updates persistent node and network models with calibration rate $\alpha=0.25$. Mean prediction error is averaged over the adaptation period following each CPU, network, or mixed change. \emph{Recovery observations} records the first point at which two consecutive relative prediction errors fall below 20\%. Results are means over ten runs; error bars are 95\% confidence intervals.

Over the adaptation periods following CPU, network, and mixed changes, Fig.~\ref{fig:e2}(a) reports mean errors of 79.1\% for Static Twin, 23.9\% for Sliding Window, and 20.3\% for Darpan. Darpan reduces prediction error by 74.3\% relative to Static Twin and by a further 15.4\% relative to Sliding Window. Static Twin does not satisfy the recovery criterion within a 16-execution phase. As Fig.~\ref{fig:e2}(b) shows, Sliding Window requires 7.5, 8.0, and 7.0 observations in the CPU, Network, and Mixed conditions, whereas Darpan requires 6.1, 7.0, and 5.9. Persistent online model updates therefore translate new physical evidence into a more accurate executable twin.

\subsection{E3: Snapshot-Based Placement Exploration}
\label{sec:e3}

\expref{3} evaluates whether a snapshot and isolated candidate sessions support placement selection. We run Stream Window in seven environments: Normal; CPU pressure at Cloud, Edge, or Fog; and network pressure on links between Device and Edge, Edge and Fog, or Fog and Cloud. The candidate set contains ten representative placements: four concentrate migratable tasks at Device, Edge, Fog, or Cloud; three split computation across adjacent tiers; and three alter where parallel branches aggregate their results. Every experimental repetition executes all candidates in all environments in randomized order, and the process is repeated 20 times. Physical runs determine the true best and worst candidates.

We compare Darpan Calibrated, Cross-workload Twin, Uncalibrated Twin, Static Cloud, and Random. Decision quality is measured by normalized placement regret
\begin{equation}
    R=\frac{T_{\mathrm{selected}}-T_{\mathrm{best}}}
    {T_{\mathrm{worst}}-T_{\mathrm{best}}},
    \label{eq:regret}
\end{equation}
where $R=0$ denotes selection of the true best placement.

\begin{figure}[!t]
    \centering
    \includegraphics[width=\columnwidth]{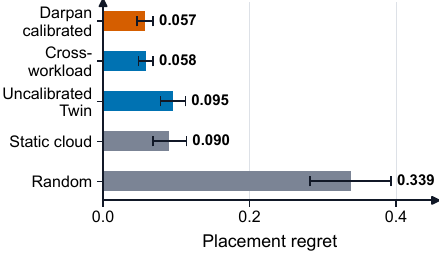}
    \caption{Normalized placement regret when candidate deployments are selected using calibrated, transferred, uncalibrated, fixed, or random information. Error bars show 95\% block-bootstrap confidence intervals over 20 paired randomized blocks.}
    \label{fig:e3}
\end{figure}

Fig.~\ref{fig:e3} reports mean regrets of 0.057 for Darpan Calibrated, 0.058 for Cross-workload Twin, 0.095 for Uncalibrated Twin, 0.090 for Static Cloud, and 0.339 for Random. Calibrated and Cross-workload therefore achieve similar average regret, but calibration selects the exact physical best in 65.0\% of cases, compared with 31.4\% for Cross-workload. Relative to Uncalibrated Twin, Static Cloud, and Random, Calibrated reduces regret by 39.9\%, 36.3\%, and 83.1\%, respectively. These results complement \expref{2}: online calibration improves exact placement choices, while transferred observations often retain enough ordering information to select a near-optimal candidate.

\subsection{E4: Failure Recovery}
\label{sec:e4}

\expref{4} evaluates the complete twin-to-physical recovery loop. CPU Analytics initially places its parallel feature-extraction tasks at Fog. Three seconds after application start, the Fog node is failed, interrupting active work. Reactive Recovery moves failed tasks to a predetermined Edge node. Darpan Recovery creates a snapshot from the latest post-failure state, compares feasible recovery nodes in the Twin, and submits the recovery action with the shortest predicted completion time. Both methods use the same initial placement and failure time and are run ten times; error bars are 95\% confidence intervals.

\begin{figure}[!t]
    \centering
    \includegraphics[width=\columnwidth]{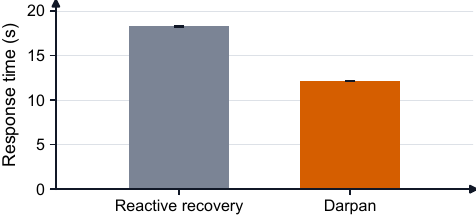}
    \caption{End-to-end response time after a Fog-node failure for fixed reactive recovery and Darpan's snapshot-guided recovery.}
    \label{fig:e4}
\end{figure}

As Fig.~\ref{fig:e4} shows, Reactive Recovery requires 18.30~s on average, whereas Darpan Recovery requires 12.18~s. Darpan reduces response time by 33.4\% and completes every DAG. The result demonstrates that snapshot-guided exploration can translate current failure observations into a faster physical recovery decision.

\subsection{E5: Execution-Plane Scale-Out}
\label{sec:e5}

\expref{5} evaluates physical execution-plane scale-out with 10, 20, 30, and 40 nodes under the same authoritative Session control plane. For each added node, we add one concurrent Scale DAG, keeping per-node offered load constant. Every scale is measured ten times. The metric is successfully completed DAGs per minute. The ideal line is a linear extrapolation from measured 10-node throughput; error bars are 95\% confidence intervals.

\begin{figure}[!t]
    \centering
    \includegraphics[width=\columnwidth]{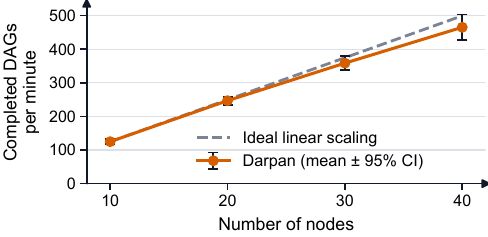}
    \caption{Measured Darpan throughput and ideal linear scaling as the physical execution plane expands from 10 to 40 nodes.}
    \label{fig:e5}
\end{figure}

As Fig.~\ref{fig:e5} shows, throughput increases from 124.9 DAGs/min at 10 nodes to 465.4 DAGs/min at 40 nodes. Scaling efficiency at 20, 30, and 40 nodes is 98.8\%, 95.7\%, and 93.1\%, respectively, so the measured curve remains close to ideal linear scaling. Each experimental round ends only after all concurrently submitted DAGs finish. With more concurrent DAGs, one slow completion has a larger effect on the round, which explains the small departure from the ideal line at 40 nodes.

\subsection{E6: Runtime Cost}
\label{sec:e6}

\expref{6} measures two operations used whenever Darpan evaluates alternatives: capturing the current state as a Snapshot and executing candidate trajectories in the Twin. We construct Session states with 4, 10, 20, and 40 nodes, increasing links, active applications, components, and observations with the node count. Ten independent runs measure Snapshot creation at each size. A separate ten-run microbenchmark measures batches of 1, 4, 8, 16, and 32 complete Wide Parallel Scheduling trajectories. We also measure the local processing needed to record one CPU-capacity observation and propagate it to the attached Twin. This test uses 24 fresh Session and Twin runs, balances every ordering of the four state sizes, and records 500 observations at each size per run.

\begin{figure*}[!t]
    \centering
    \includegraphics[width=0.94\textwidth]{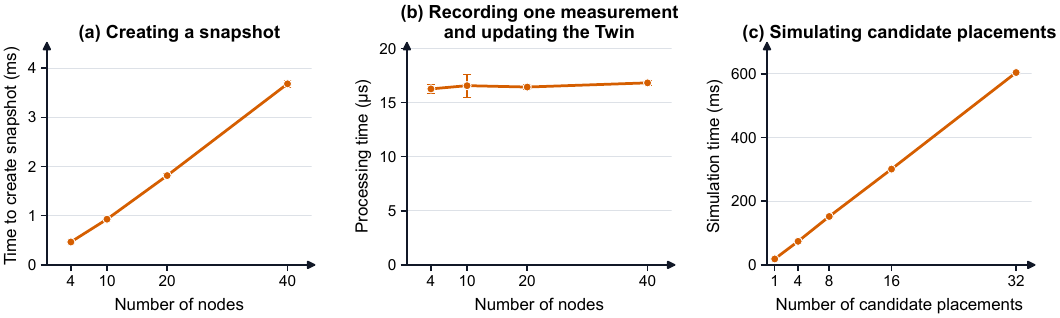}
    \caption{Darpan runtime cost. (a) Time to create a Snapshot as the live Session grows from 4 to 40 nodes. (b) Local processing time to record one CPU-capacity observation and update the attached Twin after that observation has reached the Session. (c) Wall-clock time for complete Wide Parallel Scheduling candidate batches. Error bars show 95\% confidence intervals over independent run means.}
    \label{fig:e6}
\end{figure*}

As Fig.~\ref{fig:e6}(a) shows, Snapshot creation grows from $0.466\pm0.031$~ms at 4 nodes to $3.681\pm0.071$~ms at 40 nodes. In Fig.~\ref{fig:e6}(b), recording one observation and updating the Twin averages between 16.262 and 16.821~$\mu$s across the tested state sizes; the 40-node result is $16.821\pm0.258$~$\mu$s.

Fig.~\ref{fig:e6}(c) shows that one complete candidate trajectory takes $18.718\pm0.293$~ms and a batch of 32 takes $604.337\pm4.644$~ms. The near-linear increase reflects sequential execution of progressively larger candidate batches. Together, these results show that, at the tested 40-node scale, Darpan creates an up-to-date Snapshot in a few milliseconds and updates the Twin in tens of microseconds, while candidate-execution cost grows predictably with the number of alternatives.

\subsection{E7: State Changes Between Decision and Execution}
\label{sec:e7}

\expref{7} examines a critical runtime risk: a placement can be feasible when it is selected and become infeasible before it is executed. Darpan checks the selected placement against the current physical state immediately before sending it for execution. We use a variant that checks feasibility only when selecting the placement as an ablation. This variant retains the initial check but omits revalidation against the latest recorded state at submission. Matched trials either make the selected node unavailable or consume its remaining CPU capacity. If Darpan rejects the placement, the experiment captures the current state again and submits a predefined backup placement.

\begin{figure*}[!t]
    \centering
    \includegraphics[width=0.94\textwidth]{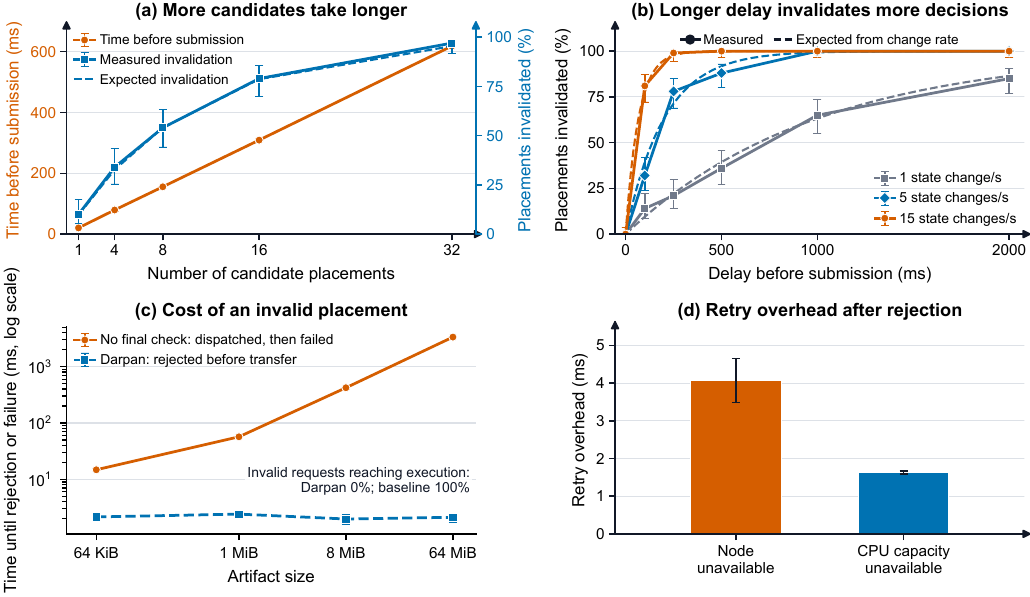}
    \caption{State changes between decision and execution. (a) Effect of candidate count on decision time and the fraction of placements invalidated before submission at a target-node failure injection rate of $\lambda=5~\mathrm{s}^{-1}$. (b) Longer delays and more frequent node changes invalidate more placements. (c) Darpan rejects invalid placements before artifact transfer, whereas the variant that validates only at selection transfers the artifact and then reports failure; the two path latencies are compared across artifact sizes. (d) Control overhead for capturing a fresh Snapshot and submitting a predefined backup after rejection. Results use ten independent runs; error bars show 95\% confidence intervals.}
    \label{fig:e7}
\end{figure*}

The experiment varies the time between selection and execution and the frequency of physical changes; every condition contains 100 paired trials across ten runs. A second experiment varies artifact size from 64~KiB to 64~MiB to measure the work wasted when an invalid placement is sent for execution. Before timing, each method completes one excluded warm-up transfer on its own otherwise identical stack, preventing first-use initialization from favoring either method.

Under a controlled target-node failure injection rate of $\lambda=5~\mathrm{s}^{-1}$, Fig.~\ref{fig:e7}(a) shows the relationship between candidate count and decision timeliness. Increasing the candidate count from 1 to 32 raises decision time from 20.44 to 618.32~ms and the fraction invalidated before submission from 10\% to 97\%. Fig.~\ref{fig:e7}(b) further shows that longer submission delays and higher state-change rates increase the likelihood of invalidation. These results quantify the timeliness cost of broader candidate exploration under the tested workload and change conditions, and show why the selected placement is checked against the current state after candidate comparison.

In the experiments, node unavailability or insufficient CPU capacity is recorded in the system state before submission. Darpan uses this state to reject every invalid placement before physical dispatch. Fig.~\ref{fig:e7}(c) further compares the cost of the two validation paths under node unavailability. Darpan takes about 2~ms on average from state capture, i.e., Snapshot creation, to rejection of the invalid placement, without transferring the artifact. The variant that validates only at selection first transfers the complete artifact and then reports failure; its mean time from state capture to failure increases from 14.792~ms for 64~KiB to 3.345~s for 64~MiB. These results show that checking at submission avoids wasted transfer and failure-handling work that grows with artifact size.

After a rejection, Darpan captures a fresh Snapshot and submits a predefined backup placement. The mean control overhead in Fig.~\ref{fig:e7}(d) is 4.066~ms after node unavailability and 1.628~ms after CPU-capacity loss. Thus, once a feasible backup is available, Darpan can submit it with only millisecond-scale control overhead.

\subsection{Case Study: Improving Online DRL Scheduling}
\label{sec:case-study}

This \casestudyref{} asks two questions: can Darpan incorporate an external scheduler through its common interface, and can Darpan-generated experience improve a scheduler that otherwise performs poorly? We run Wide Parallel Scheduling and compare all methods over the same sequence of 100 physical DAG executions and ten random initializations.

\begin{figure*}[!t]
    \centering
    \includegraphics[width=0.96\textwidth]{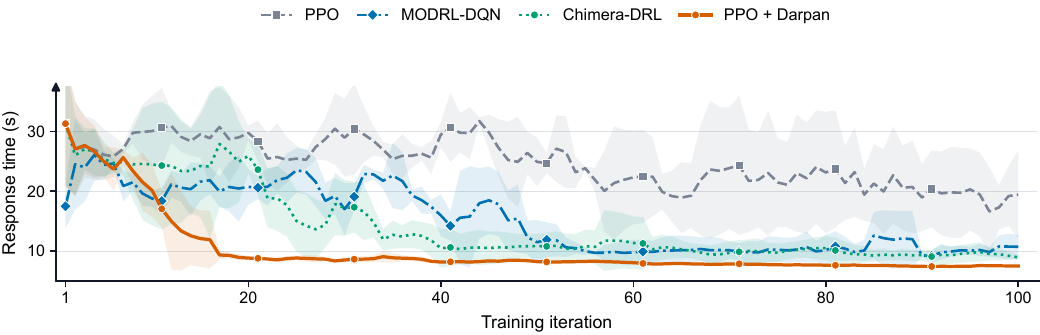}
    \caption{Case Study: physical DAG response time during online scheduler training. Curves show the mean of ten independent runs and shaded regions show one standard deviation.}
    \label{fig:case-study}
\end{figure*}

The four online methods are PPO, the state-of-the-art MODRL-DQN scheduler~\cite{xu2026online}, the state-of-the-art Chimera-DRL scheduler~\cite{long2026overselling}, and PPO+Darpan. PPO+Darpan uses the same PPO learner and physical workload sequence as PPO, with additional execution experience generated by Darpan after each physical run. The comparison therefore tests whether Darpan's experience improves the same learner.

Fig.~\ref{fig:case-study} shows that PPO and PPO+Darpan start from the same response time and diverge after Darpan-generated experience becomes available. Over the final 20 iterations, PPO, MODRL-DQN, Chimera-DRL, and PPO+Darpan average 20.19, 10.63, 9.44, and 7.53~s, respectively. PPO+Darpan reduces response time by 62.7\% relative to PPO and outperforms the two state-of-the-art schedulers by 29.1\% and 20.2\%. The \casestudyref{} therefore demonstrates both external-scheduler integration and improvement of the original learner.

\section{Future Directions}
\label{sec:vision}

\textbf{Real-world deployment and application integration.}
An important vision for Darpan is to move Digital Twins beyond isolated simulation environments and connect them directly to a running computing continuum. Researchers and system developers can deploy Darpan across real Device, Edge, Fog, and Cloud resources and integrate their own applications, services, and workloads without redesigning them around the framework. Applications from different domains, such as data analytics, intelligent transportation, healthcare, and industrial IoT, can enter the same Physical--Digital execution process through common application descriptions and execution interfaces. In this way, Darpan can evolve into a general runtime foundation through which users observe the execution of real applications and construct executable Digital Twins that remain synchronized with their physical counterparts.

\textbf{Decision methods and intelligent control.}
Once applications and infrastructure are connected, researchers can integrate their own resource-management methods into Darpan without modifying its core runtime. These methods may address placement, scheduling, migration, scaling, routing, and failure recovery, and may use mathematical optimization, heuristic search, reinforcement learning, multi-agent learning, or other forms of intelligent control. Darpan does not prescribe how a decision should be made. Instead, it provides different decision methods with a common physical state, digital execution environment, and physical execution interface. Researchers can therefore explore alternative decisions from the same running state and compare their outcomes under common execution semantics. As the \casestudyref{} illustrates, Digital Twin execution can also provide additional experience for learning-based controllers, allowing new AI-based resource-management methods to reduce costly experimentation on physical infrastructure while continuing to learn from observations produced by the real system.

\textbf{Scenario exploration and system planning.}
Beyond deciding what action should be taken at the current moment, Darpan can support the exploration of situations that have not yet occurred or that are difficult, expensive, or disruptive to reproduce physically. Researchers can introduce node failures, network degradation, resource contention, sudden workload changes, or combinations of these conditions within Digital Twin branches and study how systems respond. The same capability can also support longer-term questions about system evolution, such as adding new Edge resources, changing connectivity between Cloud and Edge environments, upgrading computing capacity, introducing new hardware, or evaluating alternative deployment structures. Digital Twin execution can therefore extend from immediate runtime decision support to resilience analysis, capacity planning, infrastructure design, and system evolution, allowing system designers to examine the possible consequences of a change before modifying the physical infrastructure.

\textbf{Open research ecosystem.}
In the longer term, we envision Darpan as an open experimentation ecosystem for computing-continuum research rather than a runtime with a fixed set of functions. Users can contribute their own applications and workloads, researchers can integrate new Controllers and Digital Twin models, and system developers can connect additional Cloud, Edge, Fog, and Device execution backends. The community can also share scenarios, failure models, experiment configurations, and benchmarks. Such an ecosystem could allow a scheduling method developed by one research group to be evaluated on the physical infrastructure of another, enable a new application to reuse existing decision methods, and support comparison of different Digital Twin models under common physical execution conditions. Further research on larger and multi-domain Digital Twins, more efficient candidate exploration, safe execution of complex operations, and trusted collaboration across heterogeneous infrastructures can broaden the range of systems that Darpan can support. More broadly, this vision is intended to encourage a computing-continuum research model in which applications, decision methods, scenarios, Digital Twin models, and physical execution experience can be shared and reproduced across research environments.

\section{Conclusions}
\label{sec:conclusion}

We proposed Darpan, a Digital Twin framework for the next-generation computing continuum. It turns live continuum state into an executable basis for exploring alternative decisions. Physical observations update its virtual counterpart, one captured state gives candidates a common starting point, and the selected decision is rechecked before physical execution. Experiments show that Darpan predicts real DAG behavior, adapts to changing conditions, improves placement and recovery, and retains 93.1\% scale-out efficiency at 40 nodes. In the tested state-change scenarios, Darpan takes about 2~ms on average from state capture to rejection of an invalidated decision and stops the request before artifact transfer, avoiding unnecessary physical work that grows with artifact size. Under the same physical-DAG execution budget, Darpan-generated experience lifts a PPO scheduler that otherwise performs poorly above two state-of-the-art DRL schedulers, demonstrating the framework's ability to enhance an external learner. These results establish the foundation for the broader vision of Darpan as an open experimentation platform where researchers can connect real continuum infrastructure, integrate their own applications and decision methods, and explore runtime decisions, failures, and future system configurations through synchronized Digital Twins before applying changes to the physical system.

\textbf{Software Availability:} The source code, documentation, and examples for Darpan are publicly available at \url{https://github.com/Cloudslab/Darpan}.

\bibliographystyle{elsarticle-num}

\bibliography{Darpan}

@inproceedings{deng2021fogbus2,
  author    = {Qifan Deng and Mohammad Goudarzi and Rajkumar Buyya},
  title     = {{FogBus2}: A Lightweight and Distributed Container-Based Framework for Integration of {IoT}-Enabled Systems with Edge and Cloud Computing},
  booktitle = {Proceedings of the International Workshop on Big Data in Emergent Distributed Environments},
  pages     = {1--8},
  year      = {2021},
  doi       = {10.1145/3460866.3461768}
}

@article{wang2025reinfog,
  author  = {Zhiyu Wang and Mohammad Goudarzi and Rajkumar Buyya},
  title   = {{ReinFog}: A Deep Reinforcement Learning Empowered Framework for Resource Management in Edge and Cloud Computing Environments},
  journal = {Journal of Network and Computer Applications},
  volume  = {242},
  pages   = {104250},
  year    = {2025},
  doi     = {10.1016/j.jnca.2025.104250}
}

@inproceedings{rosendo2020e2clab,
  author    = {Daniel Rosendo and Pedro Silva and Matthieu Simonin and Alexandru Costan and Gabriel Antoniu},
  title     = {{E2Clab}: Exploring the Computing Continuum through Repeatable, Replicable and Reproducible Edge-to-Cloud Experiments},
  booktitle = {Proceedings of the IEEE International Conference on Cluster Computing},
  pages     = {176--186},
  year      = {2020},
  doi       = {10.1109/CLUSTER49012.2020.00028}
}

@article{calheiros2011cloudsim,
  author  = {Rodrigo N. Calheiros and Rajiv Ranjan and Anton Beloglazov and C{\'e}sar A. F. De Rose and Rajkumar Buyya},
  title   = {{CloudSim}: A Toolkit for Modeling and Simulation of Cloud Computing Environments and Evaluation of Resource Provisioning Algorithms},
  journal = {Software: Practice and Experience},
  volume  = {41},
  number  = {1},
  pages   = {23--50},
  year    = {2011},
  doi     = {10.1002/spe.995}
}

@article{sonmez2018edgecloudsim,
  author  = {Cagatay Sonmez and Atay Ozgovde and Cem Ersoy},
  title   = {{EdgeCloudSim}: An Environment for Performance Evaluation of Edge Computing Systems},
  journal = {Transactions on Emerging Telecommunications Technologies},
  volume  = {29},
  number  = {11},
  pages   = {e3493},
  year    = {2018},
  doi     = {10.1002/ett.3493}
}

@article{mahmud2022ifogsim2,
  author  = {Redowan Mahmud and Samodha Pallewatta and Mohammad Goudarzi and Rajkumar Buyya},
  title   = {{iFogSim2}: An Extended {iFogSim} Simulator for Mobility, Clustering, and Microservice Management in Edge and Fog Computing Environments},
  journal = {Journal of Systems and Software},
  volume  = {190},
  pages   = {111351},
  year    = {2022},
  doi     = {10.1016/j.jss.2022.111351}
}

@article{mechalikh2021pureedgesim,
  author  = {Charafeddine Mechalikh and Hajer Taktak and Faouzi Moussa},
  title   = {{PureEdgeSim}: A Simulation Framework for Performance Evaluation of Cloud, Edge and Mist Computing Environments},
  journal = {Computer Science and Information Systems},
  volume  = {18},
  number  = {1},
  pages   = {43--66},
  year    = {2021},
  doi     = {10.2298/CSIS200301042M}
}

@article{massa2026eclypse,
  author  = {Jacopo Massa and Valerio De Caro and Stefano Forti and Patrizio Dazzi and Davide Bacciu and Antonio Brogi},
  title   = {{ECLYPSE}: A Python Framework for Simulation and Emulation of the Cloud-Edge Continuum},
  journal = {Journal of Software: Evolution and Process},
  volume  = {38},
  number  = {1},
  pages   = {e70081},
  year    = {2026},
  doi     = {10.1002/smr.70081}
}

@article{xu2026online,
  author  = {Yueshen Xu and Fanhao Zeng and Qingshan Li and Xinkui Zhao and Wei Shao and Shuiguang Deng and Rui Li},
  title   = {Online Microservice Deployment in Edge Networks via Multiobjective Deep Reinforcement Learning},
  journal = {IEEE Transactions on Services Computing},
  volume  = {19},
  number  = {3},
  pages   = {2329--2342},
  year    = {2026},
  doi     = {10.1109/TSC.2026.3683756}
}

@article{long2026overselling,
  author  = {Saiqin Long and Jianghua Qian and Jianhui Wang and Young-June Choi and Zhetao Li},
  title   = {Overselling-Aware Task Scheduling in {MEC} with Deep Reinforcement Learning},
  journal = {IEEE Transactions on Mobile Computing},
  pages   = {1--17},
  year    = {2026},
  doi     = {10.1109/TMC.2026.3709049}
}

@article{bittencourt2025continuum,
  author  = {Luiz F. Bittencourt and Roberto Rodrigues-Filho and Josef Spillner and Filip De Turck and Jos{\'e} Santos and Nelson L. S. da Fonseca and Omer Rana and Manish Parashar and Ian Foster},
  title   = {The Computing Continuum: Past, Present, and Future},
  journal = {Computer Science Review},
  volume  = {58},
  pages   = {100782},
  year    = {2025},
  doi     = {10.1016/j.cosrev.2025.100782}
}

@article{wang2024drlis,
  author  = {Zhiyu Wang and Mohammad Goudarzi and Mingming Gong and Rajkumar Buyya},
  title   = {Deep Reinforcement Learning-Based Scheduling for Optimizing System Load and Response Time in Edge and Fog Computing Environments},
  journal = {Future Generation Computer Systems},
  volume  = {152},
  pages   = {55--69},
  year    = {2024},
  doi     = {10.1016/j.future.2023.10.012}
}

@article{wang2025tfddrl,
  author  = {Zhiyu Wang and Mohammad Goudarzi and Rajkumar Buyya},
  title   = {{TF-DDRL}: A Transformer-Enhanced Distributed {DRL} Technique for Scheduling {IoT} Applications in Edge and Cloud Computing Environments},
  journal = {IEEE Transactions on Services Computing},
  volume  = {18},
  number  = {2},
  pages   = {1039--1053},
  year    = {2025},
  doi     = {10.1109/TSC.2025.3528346}
}

@article{qin2024dtnsurvey,
  author  = {Baolin Qin and Heng Pan and Yueyue Dai and Xueming Si and Xiaoyan Huang and Chau Yuen and Yan Zhang},
  title   = {Machine and Deep Learning for Digital Twin Networks: A Survey},
  journal = {IEEE Internet of Things Journal},
  volume  = {11},
  number  = {21},
  pages   = {34694--34716},
  year    = {2024},
  doi     = {10.1109/JIOT.2024.3416733}
}

@article{li2026dtorchestration,
  author  = {Tianyu Li and Xingwei Wang and Rongfei Zeng and Zhi Liu and Yuxin Zhang and Qiang He and Liang Zhao},
  title   = {A Digital Twin-Enhanced Cloud-Edge-End Collaboration Scheme for Intelligent Resource Orchestration},
  journal = {IEEE Transactions on Services Computing},
  volume  = {19},
  number  = {3},
  pages   = {1875--1890},
  year    = {2026},
  doi     = {10.1109/TSC.2026.3689495}
}

@article{saxena2025selfhealing,
  author  = {Deepika Saxena and Ashutosh Kumar Singh},
  title   = {A Self-Healing and Fault-Tolerant Cloud-Based Digital Twin Processing Management Model},
  journal = {IEEE Transactions on Industrial Informatics},
  volume  = {21},
  number  = {5},
  pages   = {4233--4242},
  year    = {2025},
  doi     = {10.1109/TII.2025.3540498}
}

@article{gao2026cost,
  author  = {Zheng Gao and Danfeng Sun and Jianyong Zhao and Huifeng Wu and Jia Wu},
  title   = {Cost-Minimized Data Edge Access Model for Digital Twin Using Cloud-Edge Collaboration},
  journal = {IEEE Transactions on Network and Service Management},
  volume  = {23},
  pages   = {252--266},
  year    = {2026},
  doi     = {10.1109/TNSM.2025.3621548}
}

@article{sedghani2025placement,
  author  = {Hamta Sedghani and Mauro Passacantando and Danilo Ardagna},
  title   = {Application Component Placement and Resource Optimization in Computing Continua},
  journal = {IEEE Transactions on Services Computing},
  volume  = {18},
  number  = {6},
  pages   = {3491--3508},
  year    = {2025},
  doi     = {10.1109/TSC.2025.3625265}
}

@article{zhao2025socialwelfare,
  author  = {Hailiang Zhao and Ziqi Wang and Guanjie Cheng and Wenzhuo Qian and Peng Chen and Jianwei Yin and Schahram Dustdar and Shuiguang Deng},
  title   = {Online Workload Scheduling for Social Welfare Maximization in the Computing Continuum},
  journal = {IEEE Transactions on Services Computing},
  volume  = {18},
  number  = {4},
  pages   = {2267--2280},
  year    = {2025},
  doi     = {10.1109/TSC.2025.3570845}
}

@article{liu2026dataorchestration,
  author  = {Yuxin Liu and Ziyi He and Xinyi Xie and Anfeng Liu and Zhetao Li and Qingyong Deng},
  title   = {Data Orchestration Service Placement and Resource Allocation Scheme for Cloud-Edge System},
  journal = {IEEE Transactions on Services Computing},
  volume  = {19},
  number  = {2},
  pages   = {1048--1061},
  year    = {2026},
  doi     = {10.1109/TSC.2026.3660225}
}

@article{ren2026joint,
  author  = {Yin Ren and Suhao Yu and Aihuang Guo},
  title   = {Joint Scheduling Mechanism for Dynamic Slice Resource Allocation and Task Offloading in User-Edge-Cloud Systems},
  journal = {IEEE Transactions on Cloud Computing},
  volume  = {14},
  number  = {1},
  pages   = {276--292},
  year    = {2026},
  doi     = {10.1109/TCC.2026.3653217}
}

@article{xia2026streamingdt,
  author  = {Qiufen Xia and Peichen Liu and Zichuan Xu and Jiankang Ren and Weifa Liang and Guangyuan Xu and Wenzheng Xu and Pan Zhou and Hao Li and Zhen Feng},
  title   = {Enabling Streaming Analytics for Digital Twin Applications in Mobile Edge Computing Networks},
  journal = {IEEE Transactions on Parallel and Distributed Systems},
  volume  = {37},
  number  = {6},
  pages   = {1295--1311},
  year    = {2026},
  doi     = {10.1109/TPDS.2026.3673833}
}

@article{chen2025medical,
  author  = {Yishan Chen and Xiangwei Zeng and Huashuai Cai and Qing Xu and Zhiquan Liu},
  title   = {Decentralized {QoS}-Aware Model Inference Using Federated Split Learning for Cloud-Edge Medical Detection},
  journal = {IEEE Transactions on Parallel and Distributed Systems},
  volume  = {36},
  number  = {10},
  pages   = {2119--2136},
  year    = {2025},
  doi     = {10.1109/TPDS.2025.3594694}
}

@article{cao2025vehicular,
  author  = {Dun Cao and Shirui Huang and Ning Gu and Fayez H. Alqahtani and R. Simon Sherratt and Jin Wang},
  title   = {Co-Optimization of Partial Offloading and Resource Allocation for Multi-User Tasks in Vehicular Edge Networks},
  journal = {IEEE Transactions on Parallel and Distributed Systems},
  volume  = {36},
  number  = {12},
  pages   = {2537--2548},
  year    = {2025},
  doi     = {10.1109/TPDS.2025.3571470}
}

@article{yang2025adaptivefl,
  author  = {Kechang Yang and Biao Hu and Mingguo Zhao},
  title   = {Coordinating Computational Capacity for Adaptive Federated Learning in Heterogeneous Edge Computing Systems},
  journal = {IEEE Transactions on Parallel and Distributed Systems},
  volume  = {36},
  pages   = {1509--1523},
  year    = {2025},
  doi     = {10.1109/TPDS.2025.3574718}
}

@article{clousim7g,
author = {Andreoli, Remo and Zhao, Jie and Cucinotta, Tommaso and Buyya, Rajkumar},
title = {CloudSim 7G: An Integrated Toolkit for Modeling and Simulation of Future Generation Cloud Computing Environments},
journal = {Software: Practice and Experience},
volume = {55},
number = {6},
pages = {1041-1058},
year = {2025}
}

\end{document}